\documentclass[amssymb, nofootinbib, superscriptaddress]{revtex4}

\usepackage{amsmath}
\usepackage{amsfonts}
\usepackage{graphicx}
\usepackage{psfrag}
\usepackage{array}
\usepackage{bbm}

\newcommand{\f}{\frac}
\newcommand{\tr}{\mathrm{tr}}

\newcommand{\su}{\mathfrak{su}}
\newcommand{\SU}{\mathrm{SU}}
\newcommand{\DSU}{\mathrm{DSU}}
\newcommand{\U}{\mathrm{U}}
\newcommand{\ISU}{\mathrm{ISU}}
\newcommand{\SO}{\mathrm{SO}}

\newcommand{\R}{\mathbb{R}}

\newcommand{\N}{\mathbb{N}}
\newcommand{\Z}{\mathbb{Z}}

\newcommand{\cF}{{\cal F}}

\newcommand{\cS}{{\cal S}}
\newcommand{\cV}{{\cal V}}

\newcommand{\id}{\mathbb{I}}

\newcommand{\be}{\begin{equation}}
\newcommand{\ee}{\end{equation}}
\newcommand{\bes}{\begin{eqnarray}}
\newcommand{\ees}{\end{eqnarray}}

\def\tl{\widetilde}

\def\arr{\rightarrow}
\def\eps{\epsilon}
\def\pp{\partial}
\def\ka{\kappa}

\def\la{\langle}
\def\ra{\rangle}
\def\vphi{\varphi}
\renewcommand{\hat}{\widehat}
\def\kk{{\cal K}}
\def\ff{{\cal F}}

\newcommand{\Ref}[1]{(\ref{#1})}
\def\nn{\nonumber}

\begin{document}
%
\title{\large\bf Matrix Models as Non-commutative Field Theories on $\R^3$}

\author{Etera R. Livine}\email{etera.livine@ens-lyon.fr}
\affiliation{Laboratoire de Physique, ENS Lyon, CNRS UMR 5672, 46 All\'ee d'Italie, 69007 Lyon, France}

\date{\small November 10, 2008}


\begin{abstract}
In the context of spin foam models for quantum gravity, group field theories are a useful tool allowing on the one hand a non-perturbative formulation of the partition function and on the other hand admitting an interpretation as generalized matrix models.
Focusing on 2d group field theories, we review their explicit relation to matrix models
and show their link to a class of non-commutative field theories invariant under a quantum deformed 3d Poincar\'e symmetry. This provides a simple relation between matrix models and non-commutative
geometry. Moreover, we review the derivation of effective 2d group field theories with non-trivial
propagators from Boulatov's group field theory for 3d quantum gravity. Besides the fact that this
gives a simple and direct derivation of non-commutative field theories for the matter dynamics
coupled to (3d) quantum gravity, these effective field theories can be expressed as multi-matrix
models with non-trivial coupling between matrices of different sizes. It should be interesting to
analyze this new class of theories, both from the point of view of matrix models as integrable
systems and for the study of non-commutative field theories.
\end{abstract}

\maketitle

\section{Introduction}
\label{intro}

Spinfoam models provide a discrete path integral formalism for (loop) quantum gravity. The theory
defines probability amplitudes for triangulated manifolds, which lead to transition amplitudes and
correlation functions between spin network states of quantum geometry. Then the partition function
is given by a sum over triangulations, which can be interpreted as a discrete version of the
Misner-Hawking sum-over-geometries. This sum can be defined non-perturbatively by a group field
theory (GFT). GFTs are generalized matrix models: they generate triangulated manifolds as Feynman
diagrams and their perturbative expansion reproduces the corresponding spinfoam amplitudes. They
are a higher dimensional extension of the standard matrix models used in (0-dimensional) string
theory which generates 2d triangulated manifolds.

The group field theory for 2d quantum gravity was studied in \cite{2d}and shown to reproduce the
known quantization of the 2d theory. In the 3d case, the spinfoam quantization of 3d quantum
gravity is given by the Ponzano-Regge state-sum model and the corresponding group field theory was
given by Boulatov \cite{boulatov}. Finally, in four space-time dimensions, it was shown that any
spinfoam model can be generated by the relevant group field theory \cite{gft}.

Besides a proposal \cite{laurent} for loop quantum gravity's physical inner product using the tree
level of GFT, group field theories are usually considered as auxiliary field theories allowing to
rigorously define the perturbative expansion of spinfoam amplitudes as sums over triangulations.
Nevertheless, some recent developments gave a non-perturbative meaning to the group field theories:
we showed that it is possible to derive from the 3d GFT some effective non-commutative (quantum)
field theories describing the matter dynamics coupled to the quantum geometry \cite{phase}. This
came as a confirmation of earlier work which proved that a certain class of spin foam observables
reproduces the evaluation of Feynman diagrams of an non-commutative field theory \cite{PR3,effqg}.
This procedure turns out to also apply to the four-dimensional case where we can derive effective
field theories with a (quantum) deformed Poincar\'e invariance starting from the 4d GFT for BF
theory \cite{4dgftdsr}. This allows a clean and clear derivation of non-commutative geometry with a
quantum deformed symmetry from non-perturbative quantum gravity.

Following this logic, we would like to propose a generic map between matrix models and non-commutative field theories on $\R^3$ (and possibly higher dimensions).
As a first step, we review the relation between the standard 2d GFT and the standard one-matrix
models. We focus on the $\SU(2)$ Lie group, but the analysis would hold from any semi-simple
(compact) Lie group. We show how to derive 2d GFT's with non-trivial kinetic terms from
Boulatov's 3d GFT following \cite{phase}. Then the key point is the introduction of the group Fourier transform which maps 2d GFT's to non-commutative field theories on a non-commutative $\R^3$ space provided with a quantum deformed Poincar\'e symmetry. This also maps the one-matrix models to field theories localized on the non-commutative sphere. We compare this case to the well-known fuzzy sphere and show there is an isomorphism between the two constructions. We study the symmetries of these field theories, discussing how the work on matrix models gives a new perspective on group field theories and vice-versa. Finally, we introduce a generalized class of matrix models, with a non-trivial coupling between matrix sizes, which are invariant under the 3d $\kappa$-deformed Poincar\'e group (more exactly the quantum double of $\SU(2)$~\footnotemark).

\footnotetext{
The reader interested in the mathematical differences between the $\kappa$-Poincar\'e symmetry and
the Drinfled double and their relevance to 3d quantum gravity can refer to \cite{catherine}. }

Most of the mathematical framework presented here is not new but the aim is to put all the pieces
together to present in a consistent way the explicit relation between group field theories for
spinfoam models, non-commutative field theories and matrix models.

\section{Group Field Theory and Non-Commutative Field Theory}

\subsection{2d Group Field Theory}

We consider a field $\vphi(g_1,g_2)$ on $\SU(2)$ which satisfies the following gauge invariance:
$$
\vphi(g_1g,g_2g)=\vphi(g_1,g_2), \quad \forall g\in\SU(2).
$$
Then the action of the two-dimensional group field theory is:
\be
S_{2d}[\vphi]\,=\,
\f{1}{2}\int dg_1dg_2\, \vphi(g_1,g_2){\vphi}(g_2,g_1)
+\f{\lambda}{3!}\int [dg]^3\, \vphi(g_1,g_2)\vphi(g_2,g_3)\vphi(g_3,g_1),
\ee
where $\lambda$ is the GFT coupling constant. Its Feynman diagrams are identified to
two-dimensional triangulations: the interaction vertex represents a (quantum) triangle and the
trivial propagator allows to glue these triangles to each other (for more details, see e.g
\cite{2d}). The combinatorics and the evaluations of these Feynman diagrams reproduce the structure
of the spin foam model for the two-dimensional topological BF theory with gauge group $\SU(2)$.
This theory is closely related to two-dimensional gravity \cite{2d,2dencore}.

We can introduce the gauge-fixed field which captures the whole gauge invariant information carried
by the field $\vphi$:
$$
\vphi(g_1,g_2)\,\equiv\,\phi(g_1g_2^{-1}).
$$
The GFT action then reads:
$$
S_{2d}[\vphi]\,=\,\f{1}{2}\int dg\, \phi(g)\phi(g^{-1}) +\f{\lambda}{3!}\int
[dg]^3\,\delta(g_1g_2g_3)\phi(g_1)\phi(g_2)\phi(g_3).
$$
We usually impose a reality condition on the field, $\phi(g^{-1})=\bar{\phi}(g)$ or equivalently
$\vphi(g_2,g_1)=\bar{\vphi}(g_1,g_2)$, so that the kinetic term can also be written simply as $\int \phi\bar{\phi}$.

We straightforwardly generalize this action to include all gauge-invariant polynomial couplings:
\be
S_{2d}[\phi]\,=\,
\sum_n\f{\alpha_n}{n!}\,\int [dg]^n\,\delta(g_1..g_n)\phi(g_1)..\phi(g_n).
\ee
The $n=2$ term is the quadratic kinetic term giving the propagator. All higher order polynomials
define interaction vertices identified to $n$-polygon: the $n=3$ term gives triangles, $n=4$
squares and so on.
To better probe the theory, we decompose the group field $\phi$ over the $\SU(2)$ representations:
\be
\phi(g)\,=\,
\sum_{j} \tr\left[\phi^jD^j(g)\right]\,=\,
\sum_{j,a,b}\phi^j_{ab}D^j_{ba}(g).
\ee
The spin $j\in\N/2$ labels the irreducible representations of $\SU(2)$. The indices $a,b$ label the
standard basis of the $\SU(2)$ representations with basis vectors diagonalizing  the generator
$J_z$. The trace $\tr[\cdot]$ is taken over each $j$-representation. $\phi^j_{ab}$ are the
coefficient matrices defining the Fourier transform of the field $\phi(g)$. Finally,
$D^j_{ab}(g)\equiv\la j,a|g|j,b\ra$ is the Wigner matrix representing the group element $g$ in the
$j$-representation.

The interesting operation for our purpose is the $\SU(2)$-convolution, $\phi\circ\psi (g)\,=\, \int
dh\, \phi(h)\psi(h^{-1}g)$. Its decomposition into $\SU(2)$ representations reads:
\be
\phi\circ\psi(g)\,=\,\sum_j \f{1}{d_j}\tr\left[\psi^j\phi^jD^j(g)\right],
\ee
where $d_j=(2j+1)$ is the dimension of the $\SU(2)$-representation of spin $j$. This allows to give
a simple expression of each interaction term of the group field:
\be
\int [dg]^n\,\delta(g_1..g_n)\phi(g_1)..\phi(g_n)\,=\,
\phi\circ..\circ\phi(\id)\,=\,
\sum_j \f{1}{d_j^{n-1}}\,\tr[(\phi^j)^n].
\ee
Introducing the renormalized matrices $M_j\,\equiv\, \phi^j/d_j$ of size $d_j\times d_j$, the 2d
group field theory can thus be expressed as a tower of decoupled matrix models\cite{2d}:
\be
S_{2d}[\phi]\,=\,
\sum_j d_j \left[\sum_n\f{\alpha_n}{n!}\tr (M_j)^n \right].
\ee
Imposing the reality condition $\phi(g^{-1})=\bar{\phi}(g)$ on the group field amounts to requiring
the Hermicity  of the matrices $(M_j)\dag=M_j$. The only relation between these matrix models of
different sizes are the coupling constants $\alpha_n$. Finally, if we restrict ourselves to fields
only exciting a specific spin $j$ representation, then we get a single matrix model of size
$(2j+1)\times(2j+1)$.

We conclude this introductory section with the remark that all the previous structures work exactly
the same way for any other compact semi-simple Lie group. The only modification is the labeling of
the irreducible representations of the group, which enters the (Peter-Weyl) decomposition of $L^2$
functions over the group. In particular, representations of a higher rank group will be labeled by
several (half-)integers instead of a single spin $j$.

\subsection{Group Fourier Transform and Star-Product on $\R^3$}

Following the previous work on the relation between (3d) spin foam models and non-commutative
quantum field theories, the main lesson to keep in mind is that the group field theory is the
``momentum" representation of a quantum field theory (and not its representation in coordinate
space). The non-locality of the 2d GFT interaction term simply relates to the non-locality of
quantum field theories written in momentum space. Nevertheless, the momentum space is now a curved
manifold (e.g here the $\SU(2)$ group) and its dual coordinate space becomes non-commutative.
However, as shown in \cite{PR3,effqg}, we still have a deformed Poincar\'e invariance and a
momentum conservation law.

In higher dimension, the group field $\phi$ is a function of more variables and there are two
sources of non-locality in the group field theory: still the one due to the curved momentum space
and the non-commutative structure and another one coming from the non-trivial combinatorial
structure of the theory (mimicking the gluing of geometric simplices in order to form space-time
triangulations).

The main mathematical tool to make explicit this correspondence between GFTs and non-commutative
QFTs is the group Fourier transform introduced in \cite{PR1,PR3,effqg}, further developed in
\cite{majid} and revisited more rigourously in \cite{karim}. It maps functions on the group $\SU(2)$ to
functions on the $\R^3$ space dual to its Lie algebra $\su(2)$. For any function $\phi(g)$ on
$\SU(2)$, we define its group Fourier transform as the following function on $\R^3$:
\be
\hat{\phi}(\vec{x})\,=\,
\int dg\, \phi(g) e^{\f12 \tr g x}, \qquad
\textrm{with} \quad x=\vec{x}\cdot\vec{\sigma}.
\ee
The matrices $\sigma_i$ are the (Hermitian) Pauli matrices generating the $\su(2)$ algebra,
normalized to have eigenvalues $\pm1$, i.e $(\sigma_i)^2=\id$ for $i=1,2,3$. Following
\cite{PR3,effqg}, we introduce the momentum $\vec{p}\in\R^3$ as the projection of the group element
$g$ on the Pauli matrices:
\be
\vec{p}\,\equiv\,\f{\ka}{2i}\,\tr(g\vec{\sigma}),\qquad
\f12 \,\tr g x\,=\,\f{i}{\ka}\,\vec{x}\cdot\vec{p}.
\ee
The parameter $\ka$ is the (3d) Planck mass and is introduced here for dimensional purposes. It
allows to control the ``no-gravity" limit of the theory \cite{PR3}. Using this momentum, we
parametrize the $\SU(2)$ group (in its fundamental two-dimensional representation) as:
\be
g\,=\,
\cos\theta \id \,+\, i\sin\theta \,\vec{u}\cdot\vec{\sigma}
\,=\,
\eps\sqrt{1-\f{p^2}{\ka^2}}\,\id\,+\,i\f{\vec{p}}{\ka}\cdot\vec{\sigma}.
\ee
The class angle $\theta\in[0,2\pi]$ parametrizes the equivalence classes of group elements under
conjugation (up to $\Z_2$) while $\hat{u}\in{\cal S}^2$ indicates the rotation axis of $g$. Since
we have the redundance $(\theta,\hat{u})\leftrightarrow(2\pi-\theta,-\hat{u})$, we restrict the
range of the angle to $\theta\in[0,\pi]$. Then the sign $\eps=\pm$ reflects the sign of
$\cos\theta$, that $\eps=+$ when $\theta\in [0,\pi/2]$ while $\eps=-$ when $\theta\in [\pi/2,\pi]$.
Using this parametrization, the normalized Haar measure and the group Fourier transform read:
\be
\int dg\,\phi(g)
\,=\, \f{1}{4\pi\ka^3}\sum_{\eps=\pm}\int_{|p|\le \ka}
\f{d^3\vec{p}}{\sqrt{1-\f{p^2}{\ka^2}}}\,\phi(g(\vec{p},\eps)),
\qquad
\hat{\phi}(\vec{x})\,=\,
\f{1}{4\pi\ka^3}\sum_{\eps=\pm}\int \f{d^3\vec{p}}{\sqrt{1-\f{p^2}{\ka^2}}}\,
e^{\f{i}{\ka}\vec{x}\cdot\vec{p}}\,\phi(g(\vec{p},\eps)).
\ee
We further introduce a $\star$-product between functions on $\R^3$ compatible with the group
product on $\SU(2)$:
\be
e^{\f12\tr g_1x} \star e^{\f12\tr g_2x} \,\equiv\, e^{\f12\tr g_1g_2x}, \quad\forall
g_1,g_2\in\SU(2).
\label{starproduct}
\ee
A first property of this $\star$-product is that it is dual to the convolution product on $\SU(2)$:
\be
\hat{\phi}\star\hat{\psi}(x)
\,=\, \int dg_1 dg_2\, \phi(g_1)\psi(g_2)e^{\f12\tr g_1g_2x}
\,=\, \int dg\, e^{\f12\tr gx}\,(\phi\circ\psi)(g).
\label{starconvo}
\ee
Then using the identity $\int d^3x\, \exp{\f12\tr gx}=\delta(g)+\delta(-g)$ as first shown in
\cite{PR1}, it allows to compute the integral:
\be
\int d^3x\,\hat{\phi}\star\hat{\psi}(x)
\,=\, \phi\circ\psi(\id)+\phi\circ\psi(-\id)
\,=\, \int dh\,\phi(h)\psi(h^{-1}) + \int dh\,\phi(h)\psi(-h^{-1}).
\label{z2sym}
\ee
This sum over $h$ and $-h$ reflects the fact that the group Fourier transform defined above is not
sensitive to the sign $\eps$. We have a few alternatives to address this ambiguity:
\begin{enumerate}

\item We can introduce an extra factor in the Fourier transform in order to kill the $\-id$
contribution, e.g of the type $(2+\tr(g))/4$, as proposed in \cite{jimmy}. This however changes the
properties of the transform, in particular its duality with the convolution product.

\item We could move to a four-dimensional point of view reflecting the embedding of $\SU(2)\sim {\cal S}^3$ in
$\R^4$ as proposed in \cite{majid}. But this is not relevant to the present work.

\item We can require the $\phi(g)$ field to be even, $\phi(g)=\phi(-g)$. This means that $\phi$
will effectively be a field over $\SO(3)\sim\SU(2)/\Z_2$ and its decomposition over representations
will involve only integer spins $j\in\N/2$ (i.e representations with odd dimensions).

\end{enumerate}
As explained in \cite{PR3,effqg}, a second property of the $\star$-product is that it leads to a
deformed addition of momenta~\footnotemark:
\be
\vec{p}_1\oplus\vec{p}_2 \,\equiv\,
\sqrt{1-\f{p_2^2}{\ka^2}}\,\vec{p}_1+\sqrt{1-\f{p_1^2}{\ka^2}}\,\vec{p}_2
+\f1\ka\vec{p}_1\wedge \vec{p}_2.
\ee
\footnotetext{
The correct way to write this deformed addition is to take into account the $\eps$ signs:
$$
(\vec{p}_1,\eps_1)\oplus(\vec{p}_2,\eps_2) \,\equiv\,
\left(\eps_2\sqrt{1-\f{p_2^2}{\ka^2}}\,\vec{p}_1+\eps_1\sqrt{1-\f{p_1^2}{\ka^2}}\,\vec{p}_2
+\f1\ka\vec{p}_1\wedge \vec{p}_2,\eps\right),
$$
where $\eps$ is the sign of: $$
\eps=\textrm{sign}\left(\eps_1\eps_2\sqrt{1-\f{p_1^2}{\ka^2}}\,\sqrt{1-\f{p_2^2}{\ka^2}}-\f{\vec{p}_1\cdot\vec{p}_2}{\ka^2}\right).
$$}
A last basic property of this Fourier transform is its expression in term of the Wigner
representation matrices. Assuming $\phi(g)=\sum_j \tr \phi^jD^j(g)$, we can show
that~\footnotemark:
\be
\hat{\phi}(\vec{x})\,=\,
\sum_{j\in\N/2} 2i^{-2j}\,\f{J_{d_j}(|x|)}{(|x|)}\,
\tr \left[\phi^jD^j(e^{-i\pi\hat{x}\cdot\vec{J}})\right].
\ee
\footnotetext{
We use the trick that $\hat{x}\cdot\vec{\sigma}$ can be expressed as a group element,
$i\hat{x}\cdot\vec{\sigma}=\exp(i\f\pi2\hat{x}\cdot\vec{\sigma})=\exp(i\f\pi2\hat{x}\cdot\vec{J})$.
This simplifies the calculation of the group Fourier transform:
$$
\int dg\, D^j_{ab}e^{\f12\,\tr g\vec{x}\cdot\vec{\sigma}}
\,=\,\int dg\, D^j_{ab}(g) e^{\f{|x|}{2i}\tr g\exp(i\pi\hat{x}\cdot\vec{J})}
\,=\,\int dg\, D^j_{ab}\left(ge^{-i\pi\hat{x}\cdot\vec{J}}\right) e^{\f{|x|}{2i}\tr g}
\,=\, \f{\beta_j(|x|)}{d_j}D^j_{ab}\left(e^{-i\pi\hat{x}\cdot\vec{J}}\right),
$$
where the $\beta$-coefficients are given by:
$$
e^{-i\f{r}{2}\tr g}\,=\, \sum_{k\in\N/2}\beta_k(r)\chi_k(g)
\,\qquad
\beta_k(r)=\int dg\, \chi_k(g)e^{-i\f{r}{2}\tr g}
\,=\,
\f2\pi\int_0^\pi \sin^2\theta d\theta\, \chi_k(\theta)e^{-ir\cos\theta}.
$$
These coefficients can finally be computed either by expanding $\exp(-i\f{r}{2}\chi_{1/2}(g))$ in
powers of $\chi_{1/2}(g)$ and then decomposing them in characters $\chi_k$ (see e.g \cite{bh}) or
by directly using the Bessel formula $e^{iz\cos\theta}=\sum_{n\in Z}i^{-n}J_n(z)e^{in\theta}$.}
The $J_n$'s are the Bessel functions (of the first kind), the operators $\vec{J}$ are the standard
$\su(2)$ generators and $\hat{x}\,\equiv \vec{x}/|x|$ is the normalized direction vector of
$\vec{x}$. In particular, this gives the Fourier transform of the $\SU(2)$ characters $\chi_j(g)$
(defined as the trace of the group element $g$ in the $j$-representation):
\be
\int dg\, \chi_j(g)e^{\f12\tr\, gx}
\,=\,
\f{2J_{d_j}(|x|)}{|x|} \quad\textrm{if }j\in\N \textrm{ and 0 otherwise}.
\ee

Now, starting with a field $\phi(g)$ on $\SO(3)$, $\phi(g)=\phi(-g)$, we compute the Fourier
transform of the group field theory. Using the identity \Ref{starconvo}, we get:
\be
S_{2d}[\phi]
\,=\,
\sum_n\f{\alpha_n}{n!}\,\int [dg]^n\,\delta(g_1..g_n)\phi(g_1)..\phi(g_n)
\,=\,
\f12\sum_n\f{\alpha_n}{n!}\,\int_{\R^3} d^3x\, \hat{\phi}\,{}^{\star n}(x).
\ee
The reality condition $\phi(g^{-1})=\bar{\phi}(g)$ on the group field simply translates into the
reality of the field $\hat{\phi}(x)\in\R$.
Seen the relation between the group field theory and (one-)matrix models, this formula is a bridge
between matrix models and (scalar) field theories on $\R^3$ provided with the non-commutative
product $\star$. As the parameter $\ka$ is sent to $\infty$, the non-commutative product becomes
the usual commutative product between functions on $\R^3$.

To understand the physical content of the theory, it is interesting to express the non-commutative
integral in term of standard integrals. Following \cite{effqg}, the non-commutative mass term can
be computed straightforwardly in term of the Laplacian $\Delta\equiv \pp_x^2$ on $\R^3$:
\be
\int d^3x\,\hat\phi\star\hat\phi(x)\,=\,
\f1{2\pi^2}\int d^3x\, \hat{\phi}(x)\,\sqrt{1+\Delta}\,\hat{\phi}(x).
\label{starFT}
\ee
The same type of formula also exists in the four-dimensional theory (see e.g \cite{4ddsr}). Thus,
even though the theory is considered as trivial on the non-commutative $\R^3$ space, it has a
in-built non-locality and non-trivial dynamics seen from the viewpoint of the standard commutative
$\R^3$ space. A remark is that the previous formula \Ref{starFT} does not seem to depend on the
physical parameter $\ka$, which is hidden in the field decomposition in term of the momentum
$\vec{p}$. But that's because the coordinates $\vec{x}$ are considered dimensionless: if we were to
re-establish their proper dimensionality, we would measure $x$ in $\hbar\ka^{-1}$ units. The
Feynman propagator of the theory is given by the inverse of the kinetic term:
\be
{\cal F}(\vec{x})\,=\,
\f1{\ka^3}\int_{|p|<\ka} \f{d^3\vec{p}}{\sqrt{1-\f{p^2}{\ka^2}}}\,e^{\f i\ka \vec{x}\cdot\vec{p}}
\,\propto\,
\f{J_1(|x|)}{|x|}.
\ee
It is completely regular at short distances as $|x|\arr 0$ and should be compared to the propagator
of a massive scalar in the standard Euclidean $\R^3$ space:
$$
F_m(\vec{x})\,=\,
\ka^2\int d^3\vec{p}\, \f{e^{\f i\ka \vec{x}\cdot\vec{p}}}{p^2-m^2+i\eta}
\,\propto\,
\f{e^{-i(\f{m}{\ka}-i\eta)|x|}}{|x|}.
$$

\subsection{Deformed Poincar\'e invariance}

One important feature of the 2d group field theory presented here is its invariance under a quantum
deformed Poincar\'e symmetry. The rotational part of the 3d Poincar\'e group is unmodified and acts
by conjugation on the field $\phi(g)$:
\be
\forall \Lambda\in\SU(2) ,\quad (\Lambda\rhd\phi)(g)\,\equiv\,
\phi(\Lambda^{-1} g \Lambda).
\ee
It is clear that such a map leaves any gauge invariant term invariant in the action:
$$
\int [dg]^n\,\phi(\Lambda^{-1}g_1\Lambda)..\phi(\Lambda^{-1}g_n\Lambda)\delta(g_1..g_n)
\,=\,
\int [dg]^n\,\phi(g_1)..\phi(g_n)\delta(g_1..g_n).
$$
The key point is that the action $\Lambda\rhd\cdot$ does not affect the constraint
$\delta(g_1..g_n)$ reflecting the conservation of momentum. Translation are a little bit tricker.
The action on multi-particle states is not the simple tensor product of the action on each
particle, but we have a modified co-product (dual to the modified addition of momenta). In the
momentum representation, translations act by multiplication by the plane waves:
\be
\forall x=\vec{x}\cdot\vec{\sigma},
\,(T_x\rhd\phi)(g)\,\equiv\, e^{\f12\tr g x} \phi(g).
\ee
Then we choose the action on multi-particle states consistent with the $\star$-product between
plane waves:
\bes
T_x\rhd\phi(g_1)\otimes\phi(g_2)\otimes..\otimes\phi(g_n) &
\equiv& e^{\f12\tr g_1..g_n x}\,\phi(g_1)\otimes\phi(g_2)\otimes..\otimes\phi(g_n) \\
&=& e^{\f12 \tr g_1 x}\phi(g_1)\star e^{\f12 \tr g_2 x}\phi(g_2)\star..\star e^{\f12 \tr g_n
x}\phi(g_n)\nonumber.
\ees
This structure with a deformed action of translations is the quantum double $\DSU(2)$, which is
similar to the $\kappa$-deformation of the Poincar\'e group $\ISU(2)$ (see \cite{effqg,majid,karim}
for more details). Due to the momentum conservation constraint $\delta(g_1..g_n)$, it is clear that
the translation $T_x$ for all $\vec{x}\in\R^3$ leave the group field theory action invariant.
Therefore we do have an field theory invariant under a quantum deformed Poincar\'e group.

One subtlety is the restriction to $\SO(3)$-fields, i.e fields satisfying the parity condition
$\phi(-g)=\phi(g)$. For this purpose, we need to symmetrize the translation operators in order that
they send an even field onto an even field. We introduce the absolute value of a group element:
\be
|g|\,\equiv\, g \quad\textrm{if}\quad \f12\tr(g)=\cos\theta>0 \qquad\textrm{and}\qquad
|g|\,\equiv\, -g \quad\textrm{if}\quad \f12\tr(g)=\cos\theta<0,
\ee
so that $\tr|g|=|\tr g|$ always remains positive. Moreover, it is easy to check that
$|g_1g_2|=|g_1||g_2|$. We define even translation operators:
\be
\tl{T}_x\rhd\phi(g_1)\otimes\phi(g_2)..\otimes\phi(g_n)\,
\equiv\, e^{\f12\tr |g_1..g_n| x}\,\phi(g_1)\otimes\phi(g_2)..\otimes\phi(g_n).
\ee
It is obvious that if $\phi(g)=\phi(-g)$, then the translated field $\exp(\f12 \tr |g|x)\,\phi(g)$
satisfies the same property. Moreover, if $g_1..g_n=\id$, then $|g_1..g_n|=\id$ so that the action
restricted to even fields remains invariant under the deformed Poincar\'e symmetry with the new
action of the translations.

\subsection{3d Group Field Theory and 2d Variations}

We conclude this section on 2d group field theories on how to derive a non-trivial kinetic term by
considering some phase of the 3d group field theory \cite{phase}. Indeed, up to now, we have only
considered the trivial kinetic term given by the (non-commutative) mass term $\int dg\,
\phi(g)\phi(g^{-1})$. We did show that the triviality of this mass term hides a non-locality and
that it contains a non-trivial dynamics. However, here, we will show how to obtain a non-trivial
kinetic term of the type $\int dg\, \phi(g)\kk(g)\phi(g^{-1})$ (for example, with $\kk(g)=\vec{p}^2$)
starting from the group field theory for 3d quantum gravity.

We start with Boulatov's group field theory \cite{boulatov} for the Ponzano-Regge model. We
consider a field $\psi(g_1,g_2,g_3)$ on $\SU(2)^3$ satisfying the gauge invariance condition
$\psi(g_1g,g_2g,g_3g)=\psi(g_1,g_2,g_3)$ and we define the action:
\bes
S_{3d}[\psi]&\equiv&
\f12\int [dg]^3 \psi(g_1,g_2,g_3)\psi(g_3,g_2,g_1) \nonumber\\
&&-\f{\lambda}{4!}\int [dg]^6\psi(g_1,g_2,g_3)\psi(g_3,g_4,g_5)\psi(g_5,g_2,g_6)\psi(g_6,g_4,g_1).
\ees
The interaction term represents a tetrahedron and the (trivial) propagator allows to glue these
tetrahedra along their boundary triangles, so that Feynman diagrams of Boulatov's group field
theory can be identified to three-dimensional triangulations. There is an issue about the
properties  of the field $\psi$ under permutations of its three arguments but this will be
discussed elsewhere \cite{withflo}. The reality condition on the field $\psi$ reads
$\psi(g_3,g_2,g_1)=\bar{\psi}(g_1,g_2,g_3)$. Finally, we could consider all possible interaction
terms\footnotemark corresponding to different 3d blocks, which would translate in higher gauge
invariant polynomial integral of the field $\psi$.
\footnotetext{
Actually, from the point of view of the renormalisation group flow, we need to consider all these
interaction terms in the effective action. Moreover, it was shown in \cite{borelsum} that we need
to add at least an extra ``pillow" term in order to make the group field theory partition function
Borel summable (i.e so that it has a non-perturbative meaning).}

The procedure introduced in \cite{phase} is to look at variations around non-trivial solutions to
the classical field equations associated to the action $S_{3d}$. These equations of motion are:
\be
\psi(g_3,g_2,g_1)=\f{\lambda}{3!}
\int dg_4dg_5dg_6\,\psi(g_3,g_4,g_5)\psi(g_5,g_2,g_6)\psi(g_6,g_4,g_1).
\ee
We consider a specific class of classical solutions, named ``flat" solutions:
\be
\psi^{(0)}(g_1,g_2,g_3)\,=\, \sqrt{\f{3!}{\lambda}}\int dg\,\delta(g_1g)F(g_2g)\delta(g_3g).
\ee
As shown in \cite{phase}, this ansatz give solutions as soon as $\int F^2=1$ (or F=0). There exists
other solutions \cite{withflo} but they are not relevant to the present discussion. We now define
the effective action for variations around such classical solutions~\footnotemark:
\be
S_{eff}[\phi]\,\equiv\, S_{3d}[\psi=\psi^{(0)}+\phi(g_1g_3^{-1})]-S_{3d}[\psi^{(0)}].
\ee
\footnotetext{
A subtle point here is that the evaluation of the action on these classical solutions is actually
infinite, $S_{3d}[\psi^{(0)}]=\infty$. The definition of the effective actions thus involve a
infinite renormalisation. This is due to the divergence of integral such as $\int \delta(g)^2$.
This could be solved by working on a $q$-deformation of $\SU(2)$ at root of unity.}
Such $\phi(g_1g_3^{-1})$ variations are obviously not generic field variations, but there are the
most general gauge-invariant variations which do not depend on the variable $g_2$. Since they only
depend on the two group elements $g_1$ and $g_3$, we call them ``two-dimensional" variations. It is
straightforward to compute the effective action:
\be
S_{eff}[\phi]\,=\,
\f12\int dg\,\phi(g)\kk(g)\phi(g^{-1})-\f\mu{3!}\int[dg]^3\, \phi(g_1)\phi(g_2)\phi(g_3)\delta(g_1g_2g_3)
-\f{\lambda}{4!}\int[dg]^4\, \phi(g_1)..\phi(g_4)\delta(g_1..g_4),
\ee
with the kinetic term and the 3-valent coupling given in term of $F$:
$$
\kk(g)\,=\,1-2\left(\int F\right)^2-\int dh F(h)F(hg),
\qquad
\f\mu{3!}\,=\,\sqrt{\f{\lambda}{3!}}\,\int F.
$$
One can show that the kinetic term is always positive, $\kk(g)\ge 0$. For more details, we refer
the interested reader to \cite{phase}. The trivial special case if given by $F=0$ which amounts to
simply computing $S_{3d}[\phi]$. The 3d group field theory then simply reduces to the 2d case
presented previously. Other examples are given by $F$ being the character of the representation of
spin $j$, then the 3-valent coupling vanishes $\mu=0$ and the kinetic term becomes
$\kk(g)=1-\chi_j(g)/d_j>0$.

The standard cases are given by $F=a\chi_1+b$ and $F=a\chi_{1/2}+b$ with $a,b$ arbitrary constants.
The kinetic term $\kk(g)$ is then respectively of the type $\kk(g)=\vec{p}^2/\ka^2-\sin^2\vartheta$ or
$\kk(g)=\cos(\theta)-\cos(\vartheta)$, where $\theta$ is still the class angle of the group element $g$ and $\vartheta\in[0,\pi/2]$ is an angle depending on the constants $a,b$. The first case gives exactly the standard scalar field on the
non-commutative $\R^3$ space dual to $\SO(3)$ \cite{PR3, majid, karim}. The Feynman propagator is
the inverse of $\kk(g)$ and its decomposition in $\SU(2)$ representations is particularly
simple~\cite{effqg}:
\be
\ff(g)\,=\,\f{1}{\f{p^2}{\ka^2}-\sin^2\vartheta+i\eta}\,=\,\f2{\cos\vartheta}\sum_{j\in\N}e^{-id_j(\vartheta-i\eta)}\chi_j(g).
\ee
The second case is a slight modification of the first. Since the kinetic term is given by
$\chi_{1/2}(g)$, the propagator sees the whole $\SU(2)$ structure and the field theory is not
compatible with the restriction to even fields. The Feynman propagator is very similar to the
previous one, except that it excites all the $\SU(2)$ modes $j\in\N/2$, as shown in \cite{jimmy}:
\be
\ff(g)\,=\,\f{1}{\cos\theta-\cos\vartheta-i\eta}\,=\,-\f2{\cos\vartheta}\sum_{j\in\N/2}e^{-id_j(\vartheta-i\eta)}\chi_j(g).
\ee

In the present work, we are not interested by the properties of specific examples of these
effective theories, but we focus on two aspects:
\begin{itemize}

\item All these effective theories with a non-trivial kinetic term of the type $\int dg\,
\phi(g)\kk(g)\phi(g^{-1})$ are invariant under the deformed action of the Poincar\'e group
described above.

\item They can be translated into matrix models \cite{phase} by decomposing the field $\phi$ into
$\SU(2)$ representations. However, the presence of the non-trivial factor $\kk(g)$ creates a
coupling between matrices of different sizes. The matrix models do not decouple anymore as in the
standard case given by $\kk(g)=1$. Thus this provides a non-obvious generalization of matrix
models.
Studying these multi-matrix theories would provide a new approach to these non-commutative quantum field with a deformed Poincar\'e symmetry. In particular, they bypass some of the technical and conceptual problems usually encountered in non-commutative field theories such as the braiding of Feynman diagrams, deformed canonical relations, non-trivial statistics and a deformed Fock space to account for the non-commutativity of field excitations.
In reverse, this Poincar\'e invariance could lead to non-trivial features in matrix models.

\end{itemize}

\section{Matrix Models and Fields on the Non-Commutative Sphere}

\subsection{The Fourier Transform of Matrix Models}

Let us consider a one-matrix model with the following action for a $N\times N$ (Hermitian) matrix
$M$:
$$
S[M]\,=\,\f12\tr M^2 +\f\lambda{n!}\tr M^n.
$$
Then, as shown in the previous section, it can be recasted as a group field theory by restricting
the field $\phi$ to excite a single representation. More precisely, we choose the spin $j$ such
that $N=d_j$, then:
\be
S[M]\,=\,\f{1}{d_j}\left[
\f12\int dg\, \phi(g)\phi(g^{-1}) +\f{\lambda}{n!}\int [dg]^n\,\delta(g_1..g_n)\phi(g_1)..\phi(g_n),
\right]
\ee
with the field $\phi(g)\,=\,d_j\tr MD^j(g)$ defined without any summation over the representation
label $j$. The next step is to define the group Fourier transform of the field $\phi$:
\be
\hat{\phi}(\vec{x})\,=\,
\int dg\, \phi(g)e^{\f12\tr gx}
\,=\,2d_ji^{-2j}\,\f{J_{d_j}(|x|)}{|x|}\,\tr MD^j(e^{-i\pi\hat{x}\cdot\vec{J}}).
\label{formula}
\ee
As expected, fixing a particular size $j$ amounts to fixing the radial dependence of the Fourier
field $\hat{\phi}(\vec{x})$. More precisely, $J_{d_j}(r)/r$ is peaked about $r\sim d_j$ so that the
field $\hat{\phi}$ is localized around the sphere $|x|\sim d_j$. We can also compute the average:
$$
\la r\ra =\f{\int dr\, J_{d_j}(r)}{\int dr\, J_{d_j}(r)/r}\,=\,d_j.
$$
%
The $\star$-product is then simply given by the matrix multiplication:
\be
\hat{\phi_1}\star\hat{\phi_2}\,(\vec{x})\,=\,
\int dg\, \phi_1\circ\phi_2(g)e^{\f12\tr gx},
\qquad\textrm{with}\quad
\phi_1\circ\phi_2(g)=d_j\tr M_2M_1D^j(g).
\ee

The matrix action is then easily written in term of this field on $\R^3$, at least for integer
representation $j\in\N$~\footnotemark:
\be
S[M]\,=\,\f{1}{2d_j}\left[
\f12\int d^3x\,\hat{\phi}\star\hat{\phi}(x) +\f{\lambda}{n!}\int (\hat{\phi})^{\star n}(x)
\right].
\ee
\footnotetext{
For half-integer representations, $j\in\N+\f12$, the integral $\int d^3x\,\hat{\phi}\star\hat{\phi}(x)$
vanishes as one can see from eqn.\Ref{z2sym} due to the insensitiveness of the group Fourier
transform to the sign $\eps$. As discussed in the previous section, this might be cured by a
suitable modification of the $\star$-product and the group Fourier transform.}
Now, not only the radial dependence of the field is fixed but also we are not allowed arbitrary
fields on the sphere ${\cal S}^2$ . We are allowed a finite number of modes and the angular part of the field is necessarily of the type $\tr MD^j(e^{-i\pi\hat{x}\cdot\vec{J}})$ where the trace is of course taken in the $j$-representation.
%

The natural question in such a setting is how well can one localize directions on the two-sphere?
To this purpose, we use the $\SU(2)$ coherent states (see e.g \cite{coherent}). Having fixed a
direction $\hat{y}\in{\cal S}^2$, we choose a group element $h_y$ which maps the north pole
$\hat{e}\equiv(0,0,1)$ onto $\hat{y}=h_y\rhd\hat{e}$. The standard choice (of section) is choosing
$h_y$ such that its rotation axis lays in the equatorial plane. The semi-classical state associated
to $\hat{y}$ is $|j, \hat{y}\ra\,\equiv\,h_y|j,j\ra$ where $|j,j\ra$ is the highest weight vector
of the $j$-representation (i.e the one with magnetic moment $m=j$). Then, we consider the function
on ${\cal S}^2$:
\be
f_y(\hat{x})\,\equiv\,
\la j,\hat{y}|D^j(e^{-i\pi\hat{x}\cdot\vec{J}})|j,\hat{y}\ra
\,=\,
\la j,j | D^j(e^{-i\pi (h_y^{-1}\rhd\hat{x})\cdot\vec{J}})|j,j\ra
\,=\, \tr M^{(y)}D^j(e^{-i\pi\hat{x}\cdot\vec{J}}),
\ee
with the matrix $M^{(y)}_{ab}\,\equiv\,\la j,a|j,\hat{y}\ra\la j,\hat{y}|j,b\ra$ given in term of
the decomposition of the coherent basis in the standard basis. Taking into account that the vector
$|j,j\ra$ is the $(2j)$-th tensor power of the vector $|\f12,\f12\ra$ of the fundamental
representation, we can easily compute the value of the function $f_y$:
\be
f_y(\hat{x})\,=\, \left[-i\,(h_y^{-1}\rhd\hat{x})\cdot\hat{e}\right]^{2j}
\,=\,
\left[-i\,\hat{x}\cdot\hat{y}\right]^{2j}.
\ee
$f_y$ is always smaller than 1 in modulus. It is real when $j\in\N$ and it reaches its highest
value in $\hat{x}=\pm\hat{y}$ with the sign depending on the parity of the integer $j$. This
prescription gives the best way to localize points on ${\cal S}^2$ and the precision clearly
increases with the spin $j$ i.e with the matrix size $N$.

The no-gravity limit in the 3d spin foam models is identified to the limit $\ka\arr \infty$ (the
Planck mass is sent to infinity). Here, the same way, we define the classical limit as the double
limit $\ka\arr\infty,\, j\arr\infty$ in which we recover classical abelian fields on the
two-sphere. It is possible to expand the correlations of the field theory in powers of $\ka^{-1}$
\cite{PR3} and it would be interesting to compare this semi-classical regime to the usual
``double-scaling" limit of matrix model. A priori the difference lays in the fact that the double
scaling limit involves a rescaling of the matrix coupling $\lambda$ while in the present context we rescale the momentum unit $\ka$ of the Fourier transform. Nevertheless, there might be a relation
between these two regimes of matrix models.

\subsection{Relation to the Fuzzy Sphere}

The present construction is similar to fuzzy geometries where points can not be precisely located. In particular, it would be interesting to investigate the precise relation between this model and the well-studied fuzzy sphere \cite{madore}.

The fuzzy sphere can be understood as a consistent truncation of the algebra of functions over the 2-sphere. We consider the spherical harmonics $Y^l_m(\hat{x})$, with $l\in\N$ and $-l \le m\le l$, which form an orthogonal basis of $L^2$ functions over the sphere $\cS^2$. The fuzzy sphere construction introduces a non-commutative (but associative) $\star_j$-product between the $Y^l_m$, depending on a fixed parameter $j\in\N$, such that the restriction to the sector $l\le 2j$ is stable under $\star_j$ and that this truncated product converges to the actual true product in the limit $j\arr\infty$. 

Following \cite{fuzzy}, we choose the following normalization of the spherical harmonics:
\be
Y^l_m(\hat{x})=\,i^{-m}\la l,m|h_x|l,0\ra,\quad
\bar{Y}^l_m=(-1)^mY^l_{-m},\qquad
\int_{\cS^2} \f{d^2\hat{x}}{4\pi}\,\bar{Y}^l_m Y^{l'}_{m'}=
\f{\delta^{ll'}\delta_{mm'}}{d_l},
\ee
where the group element $h_x\in\SU(2)$ maps as previously the north pole $\hat{e}\equiv(0,0,1)$ onto the relevant unit vector on the sphere $\hat{x}=h_x\rhd\hat{e}$. The usual product on the sphere gives:
\be
Y^{l_1}_{m_1}(\hat{x})\,Y^{l_2}_{m_2}(\hat{x})\,=\,
\sum_{l_3=|l_1-l_2|}^{l_1+l_2}\sum_{m_3} d_{l_3} Y^{l_3}_{m_3}(\hat{x})
C^{l_1}_{m_1}{}^{l_2}_{m_2}{}^{,l_3}_{,m_3}
\overline{C^{l_1}_{0}{}^{l_2}_{0}{}^{,l_3}_{,0}},
\ee
where $m_3=m_1+m_2$ and the $C$'s are the (normalized) Clebsh-Gordan coefficients describing the decomposition of the tensor product $V^{l_1}\otimes V^{l_2}$ into the $V^{l_3}$ irreducible representations.

Now fixing $j\in\N$, the spherical harmonics $Y^l_m$ with the restriction $l\le 2j$ span the Hilbert space $\bigoplus_{l=0}^{2j}V^l$, which is actually isomorphic to $V^j\otimes \overline{V^j}=\rm{End}(V^j)$. Exploiting this isomorphism, the fuzzy sphere can be constructed by mapping the spherical harmonics to $d_j\times d_j$ matrices in $\rm{End}(V^j)$ following \cite{fuzzy}:
\be
Y^l_m(\hat{x})
\,\longmapsto\,
[\Theta^l_m]_{ab}\,\equiv\, \sqrt{d_j}\,C^{j}_{a}{}^{j^*}_{b}{}^{,l}_{,m}
\,=\,
(-1)^{j-b}C^{j}_{a}{}^{j}_{-b}{}^{,l}_{,m},
\ee
where $j^*$ stands for the complex representation to $j$. Then the $\star_j$-product on the fuzzy sphere is defined once again simply by the matrix multiplication:
\be
Y^{l_1}_{m_1}\star_j Y^{l_2}_{m_2}
\,\longmapsto\,
\Theta^{l_1}_{m_1}\Theta^{l_2}_{m_2}
\,=\,
\sqrt{d_j}
\sum_{l_3=0}^{2j} \sum_{m_3} \Theta^{l_3}_{m_3}\, d_{l_3}C^{l_1}_{m_1}{}^{l_2}_{m_2}{}^{,l_3}_{,m_3}
\,\left\{
\begin{array}{ccc}
l_1 & l_2 & l_3 \\
j & j& j
\end{array}
\right\},
\ee
which can be expressed in term the Clebsh-Gordan coefficients and the $\{6j\}$-symbol. One can check that this converges toward the classical product when the representation $j$ is sent to infinity. Moreover, the $\star_j$-product is shown to be associative using the Biedenharn-Elliott (or pentogonal) identity on the $\{6j\}$-symbol.

The relation with the $\star$-product that we used here in rather straightforward, since both are expressed in term of the standard matrix multiplication. Thus starting from the matrices $\Theta^l_m$, we can reconstruct the corresponding fields $\hat{\phi}^l_m$ according to the formula \Ref{formula} which will multiply under our $\star$-product. Ignoring the radial components, we focus on the angular part of the fields:
\be
\hat{\phi}^l_m(\hat{x})\,=\,
\tr \,\Theta^l_m\,D^j(e^{-i\pi\hat{x}\cdot\vec{J}}).
\ee
Using the integral formula for Clebsh-Gordan coefficients~\footnotemark, we can compute explicitly this trace:
\be
\hat{\phi}^l_m(\hat{x})
\,=\,
\f{\sqrt{d_j}}{C^{j}_{j}{}^{j}_{-j}{}^{,l}_{,0}}
\int dg\, D^l_{m 0}(g) D^j_{jj}(g^{-1}e^{-i\pi\hat{x}\cdot\vec{J}}g)
\,=\,
i^m\f{\sqrt{d_j}}{C^{j}_{j}{}^{j}_{-j}{}^{,l}_{,0}}
\int_{\cS^2}\f{d^2\hat y}{4\pi}\,
Y^l_m(\hat{y})\,f_y(\hat{x}),
\ee
where $f_y(\hat{x})=(-i)^{2j}(\hat{x}\cdot\hat{y})^{2j}$ is the localizing kernel computed above.
\footnotetext{
The product of two Clebsh-Gordan coefficient is the integral over $\SU(2)$ of the product of three matrix elements. For instance, we use here~:
$$
C^{j}_{a}{}^{j^*}_{b}{}^{,l}_{,m}C^{j}_{j}{}^{j^*}_{j}{}^{,l}_{,0}
\,=\,
\int dg\, D^j_{aj}(g)\,\overline{D^j_{bj}(g)}\,\overline{D^l_{m0}(g)}.
$$
}
Thus the new functions $\hat{\phi}^l_m(\hat{x})$ are obtained from the original spherical harmonics $Y^l_m(\hat{x})$ through a simple linear transform. The kernel $f_y(\hat{x})$, which depends on the parameter $j$, allows to ``truncate" the spherical harmonic to the fuzzy sphere.

\medskip

To summarize, when we restrict the field $\hat{\phi}(x)$ to live in a single representation $j$ thus localizing it around the sphere of radius $j$, there is a simple map between the fuzzy sphere and our non-commutative sphere. Through the linear transform $Y^l_m\mapsto\Theta^l_m\mapsto\hat{\phi}^l_m$, we have mapped the fuzzy sphere $\star_j$-product to our $\star$-product dual to the convolution product on $\SU(2)$. Technically, we call $\cF$ the map computed above:
$$
(\cF[Y^l_m])(\hat{x})=\hat{\phi}^l_m(\hat{x})
\,=\,
\f{\sqrt{d_j}}{i^{2j}\,C^{j}_{j}{}^{j}_{-j}{}^{,l}_{,0}}
\int_{\cS^2}\f{d^2\hat y}{4\pi}\,
i^mY^l_m(\hat{y})\,(\hat{x}\cdot\hat{y})^{2j},
$$
then we have the equality:
\be
\cF\,\left[Y^{l_1}_{m_1}\star_j Y^{l_2}_{m_2}\right]\,=\,
\hat{\phi}^{l_1}_{m_1}\star \hat{\phi}^{l_2}_{m_2},
\ee
which generalizes to all the functions on the (fuzzy) sphere since the spherical harmonics are a basis of the space of functions.
The advantage of the $\star$-product is that it extends to a consistent non-commutative structure on the full $\R^3$, thus stacking in a consistent way all the fuzzy sphere to make the complete non-commutative $\R^3$ space.

Finally, we would like to point out that the fuzzy sphere construction was shown to be related to the Wess-Zumino-Witten model for the $\SU(2)$ gauge group \cite{WZW}. On the other hand, the  $\star$-product we discuss in this paper was shown to be related to the Ponzano-Regge model \cite{PR3,effqg}. The isomorphism between the two $\star$ and $\star_j$ products hints towards a link between the WZW theory and the spinfoam model for 3d gravity. At the classical level, such a relation is already established, since there is a clear relation between BF theory, Chern-Simons theory and the WZW model. Nevertheless, there could be a more direct connection at the quantum level.

\subsection{Mapping Matrix Models to Arbitrary Lie Groups?}

All the procedure seems to equally work with an arbitrary (compact and semi-simple) Lie group ${\cal G}$ as long as there exists an irreducible representation of that group with dimension equal to the matrix size $d=N$. We then define the group Fourier transform the same way as for $\SU(2)$ using the projection of group elements on the Lie algebra $\mathfrak{g}$ in the fundamental
representation. We would get at the end a reformulation of the matrix model in term of a field
theory in a $(\dim\mathfrak{g})$-dimensional space with the field localized around the co-adjoint
orbit corresponding to the chosen representation.

More precisely, starting with an arbitrary compact Lie group $G$, we work with a field $\phi(g)$ on the group manifold and define as above the following action:
$$
S[\phi]=\f12\int dg\, \phi(g)\kk(g)\phi(g^{-1})+\f{\lambda}{n!}\int [dg]^n\,\delta(g_1..g_n)\phi(g_1)..\phi(g_n),
$$
where $dg$ is the Haar measure on the group $G$ and the kinetic term $\kk(g)$ is assumed to be invariant under conjugation. We can define a Fourier transform between fields on the group manifold and fields on the Lie algebra $\mathfrak{g}\sim\R^\Delta$ with $\Delta=\dim\mathfrak{g}$~:
$$
\forall x\in\mathfrak{g},\quad
\hat{\phi}(x)\,=\, \int dg\, \phi(g) e^{\tr g x},
$$
where the trace is taken a priori in the fundamental representation. Choosing an orthonormal basis $b_1,..,b_\Delta$ for the vector space $\mathfrak{g}$ then defines a ``flat" momentum $\vec{p}(g)$ for each group element:
$$
p_k(g)\equiv\,-i\tr g b_k,\qquad
\tr\, g x = i\sum _k p_k x_k.
$$
Radial fields $\hat{phi}(x)$ on $\mathfrak{g}$ are those which are invariant under the action by conjugation of the group $G$, i.e. the fields which have constant values on the co-adjoint orbits.
We define a non-commutative but associative star-product between fields on $\mathfrak{g}$ by postulating a trivial composition of the plane waves dual to the convolution product on the group, $\exp(\tr xg_1)\star\exp(\tr xg_2)\equiv\exp(\tr xg_1g_2)$. This Fourier transforms together with the $\star$-product allows to map the previous group field theory to a non-commutative field theory. This field theory is moreover invariant under the quantum double $D(G)$ of the group $G$. The group $G$ acts as the Lorentz group by conjugation on the field:
$$
\forall \Lambda\in G\,(\Lambda\rhd\phi)(g)\equiv\phi(\Lambda^{-1} g \Lambda),\qquad
(\Lambda\rhd\hat{\phi})(g)\,=\,\hat{\phi}(\Lambda^{-1} x \Lambda).
$$
Then we define translations which acts by multiplication on fields $\phi(g)$,
$$
\forall x\in\mathfrak{g},\,
(T_x\rhd\phi)(g)\,\equiv\,
e^{\tr\, xg}\phi(g),
$$
with the non-trivial co-product for the action on many fields:
$$
T_x\rhd \phi_1(g_1)\phi_2(g_2)\dots
\,=\,
e^{\tr\,xg_1g_2\dots}\phi_1(g_1)\phi_2(g_2)\dots
\,=\,
(e^{\tr\,xg_1}\star e^{\tr\,xg_2}\star\dots)\,\phi_1(g_1)\phi_2(g_2)\dots
$$
We would need to check if $\int_{\mathfrak{g}} d^\Delta x\, \exp(\tr\,xg)\propto\delta(g)$. This means understanding the relation between group elements $g$ and their projection on the Lie algebra $p$. A priori, just like in the $\SU(2)$ case, it is likely that they will be a discrete symmetry which leads to one-to-many map with many group elements having the same projection. In such a case, we would have to quotient by this discrete symmetry.

If this procedure works, we have our mapping from the group field to a non-commutative field theory on $\R^\Delta$. Then decomposing the field into irreducible representations maps this same group field theory to a matrix model where the sizes of the matrix modes are given by the dimensions of the irreps of $G$. This shows another link between matrix models and non-commutative field theories on $\R^\Delta$. Finally, if we want to restrict ourself to a single matrix size $N$, then we need to identify an irrep of $G$ which has the same dimension and restrict our field $\phi(g)$ to live in that representation. Since there is a relation between co-adjoint orbits and  irreps, it looks likely that such a restricted field will have the interpretation of living on a fuzzy version of the orbit. This can be checked by computing explicitly the Fourier transform of the matrix elements of $g$. We postpone the details of such a generalization to future investigation.

\subsection{Symmetries and Diagonalization}

As we showed in the previous section, the full 2d group field theory is invariant under a deformed
action of the Poincar\'e group. If we consider a single matrix model by restricting the field to a
single representation of $\SU(2)$, this breaks this Poincar\'e invariance. More precisely, it
breaks the invariance under translations since translations mix the $\SU(2)$ representations.
Nevertheless the matrix model is still invariant under $\SU(2)$ rotations since they do not mix
representations:
\be
\phi(g)\arr \phi(\Lambda^{-1}g\Lambda),\qquad
M\arr D^j(\Lambda)MD^j(\Lambda^{-1}.)
\ee
Actually, the matrix model is invariant under the full $\U(N)$ unitary group, $M\arr UMU^{-1}$, and
the 3d rotations are simply the subgroup of matrices $U=D^j(\Lambda)$ for $\Lambda\in\SU(2)$.

First, this means that the 2d group field theory (with the trivial kinetic term) has a much larger
symmetry group than the Poincar\'e group. Indeed it is invariant under the product
$\U(1)\times\U(2)\times\U(3)\times..$. We can choose one unitary matrix $U_{(j)}$ for each
representations of $\SU(2)$ and rotate each field mode independently:
\be
\phi(g)\,=\,\sum_j \tr{}\, \phi^j\,D^j(g) \quad\arr\quad
\tl{\phi}(g)\,=\,\sum_j \tr{}\, U_{(j)}\phi^jU_{(j)}^{-1}\,D^j(g).
\ee
This symmetry holds only when the kinetic term of the group field theory is trivial, i.e equal to
$\int dg\, \phi(g)\phi(g^{-1})$ up to a constant. As soon as it becomes non-trivial with a non-constant
propagator $\kk(g)\ne 1$, this degeneracy is killed and the unitary invariance broken down back to
the invariance under 3d rotations (and, of course, the Poincar\'e group if we do not put any
restriction on the representations).

Second, a natural question is to find a geometrical meaning to these unitary transformations. Their
action on the angular part of the Fourier-transformed field is:
\be
\tr \,M\,D^j(e^{-i\pi\hat{x}\cdot\vec{J}})
\quad\arr\quad
\tr \,M\,U^{-1}D^j(e^{-i\pi\hat{x}\cdot\vec{J}})U.
\ee
When $U=D^j(\Lambda)$ is a 3d rotation, this rotates $\hat{x}$ to $\Lambda\rhd\hat{x}$. However for
a generic unitary, we can not interpret the action of $U$ as a simple transformation on $\hat{x}$.
We interpret them as generalized changes of coordinates on the two-sphere - some kind of ``fuzzy
diffeomorphisms" of ${\cal S}^2$. It is straightforward to compute the matrix elements $\la
j,a|D^j(e^{-i\pi\hat{x}\cdot\vec{J}})|j,b\ra$ by expressing the basis vectors as tensor products of
$(2j)$ vectors of the fundamental representation. This gives homogeneous polynomials of degree
$(2j)$ in the 3d coordinates of $\hat{x}$, but they do not have an obvious geometrical
interpretation. At the end of the day, acting with unitaries $U\in\U(N)$ amounts to sweeping
through all unitary-equivalent representations of $\SU(2)$ embedded in $\U(N)$.

Despite this, the diagonalization of the matrix $M$ in order to gauge fix the matrix model
partition function appears to have a geometrical interpretation. We choose the standard orthonormal
basis of the $j$-representation and restrict to diagonal modes of the type
$D^j_{mm}(e^{-i\pi\hat{x}\cdot\vec{J}})$ for $m$ running from $j$ to $-j$. They are polynomial
depending only only on the $z$ coordinate of $\hat{x}$. For instance, we compute (using the
identity $|\hat{x}|^2=1$):
\bes
\la j,j|e^{-i\pi\hat{x}\cdot\vec{J}}|j,j\ra &=& (-i)^{2j} \hat{x}_z^{2j} \nn\\
\la j,j-1|e^{-i\pi\hat{x}\cdot\vec{J}}|j,j-1\ra &=&
(-i)^{2j}\hat{x}_z^{2j-2}\left[(2j-1)-(2j)\hat{x}_z^2\right] \\
\la j,j-2|e^{-i\pi\hat{x}\cdot\vec{J}}|j,j-2\ra &=&
(-i)^{2j}\hat{x}_z^{2j-4}\left[(j-1)(2j-3)-2(j-1)(2j-1)\hat{x}_z^2+j(2j-1)\hat{x}_z^4\right]\dots\nn
\ees
It is clear that these diagonal matrix elements provide a basis for polynomials of degree $j$ in
$\hat{x}_z^2$. Diagonalizing the matrix $M$ thus amounts to choosing a direction on ${\cal S}^2$
(the ``z" direction) and considering fields which are polynomials of degree less or equal to $j$ in
the coordinate squared of the $\hat{x}$ along the chosen axis. This constraint that the field only
depends on $\hat{x}_z$ is much stronger than a straightforward gauge fixing of the 3d rotations.
Here the gauge fixing corresponding to the diagonalization leads to the reduction of the system
living on the two-sphere to a one-dimensional theory, breaking down the $\SU(2)$ symmetry down to $\U(1)$. Of course, to complete this analysis, one
should compute the Fadeev-Popov determinant. But the purpose here was to provide the field
diagonalization with a geometric interpretation in $\R^3$.


\subsection{Extended Matrix Models: Coupling Sizes}

Since the link between the 2d group field theory and the one-matrix model is explicit, we can
import methods and results from the study of matrix models to compute the partition function of the
group field theory. Following the standard calculation (see e.g \cite{matrix}), we diagonalize the
matrix $M$ and express the partition function of the $N\times N$ matrix model as:
\be
Z_N\,\equiv\,
\f{1}{\cV_N}\int [dM]\, e^{-\tr V(M)}
\,=\,\f{1}{N!}\int \prod_{k=1}^N\f{d\lambda_k}{2\pi}\,\Delta(\lambda)\,e^{-\sum_kV(\lambda_k)},
\ee
where $\Delta(\lambda)$ is the Fadeev-Popov determinant of the gauge fixing by diagonalization and
$\cV_N$ is the volume of the the unitary group $\U(N)$~:
$$
\Delta(\lambda)\,=\,\prod_{k<l}(\lambda_k-\lambda_l)^2,\qquad
\cV_N\,=\,\f{(2\pi)^{N(N+1)/2}}{\prod_{k=1}^{N-1}k!}.
$$
We introduce the orthogonal polynomial associated to the potential $V(\lambda)$:
\be
\int \f{d\lambda}{2\pi}\,e^{-V(\lambda)}\,P_n(\lambda)P_m(\lambda)=h_n\delta_{nm},
\ee
where the polynomials are normalized by requiring the behavior $P_n(\lambda)=\lambda^n+\dots$. Then
one can compute the partition function:
\be
Z_N\,=\,\prod_{n=0}^{N-1}h_n\,=\,h_0^N\prod_{n=1}^{N-1}r_n^{N-n},
\ee
where $2\pi h_0=\int d\lambda \exp(-V(\lambda))$ and the coefficients $r_n\,\equiv h_n/h_{n-1}$
enter the recursion relation for the polynomials:
$$
(\lambda+s_n)P_n(\lambda)\,=\,P_{n+1}(\lambda)+r_nP_{n-1}(\lambda).
$$
As it is well-known, for a quadratic potential $V(M)\,=\,M^2/2$, the relevant orthogonal
polynomials are the Hermite polynomials and we get:
\be
P_n(\lambda)=\f{1}{(\sqrt{2})^n}\,H_n(\f{\lambda}{\sqrt{2}}),
\qquad h_n=\f{1}{\sqrt{2\pi}}\,n!
\ee

We can use these results to compute the partition function of the 2d group field theory.
Considering the group field action,
$$
S_{2d}=\f12\int dg \phi(g)\phi(g^{-1})+\sum_{n\ge3}\f{\alpha_n}{n!}\int
[dg]^n\,\delta(g_1..g_n)\phi(g_1)..\phi(g_n),
$$
we decompose the field in $\SU(2)$ representations and express the partition function as a tower of
matrix models following the calculations of section I:
\be
Z\,=\,\int [d\phi]\, e^{-S_{2d}[\phi]}
\,=\,
\int[d\phi]\, \prod_j e^{-d_j\,\tr\, W[M_j]},
\ee
where the potential $W[M]=\f12M^2+\sum_{n\ge3}\f{\alpha_n}{n!}\,M^n$ is defined independently of
the matrix size. From the matrix model perspective, the $d_j$ factor in front of the potential is
the right one in order to look at the large matrix size regime with saddle point techniques and
study the double scaling limit. However, here we do not only look at a single matrix model in the
limit $j\arr \infty$ but we must consider the whole tower of matrix models with all possible sizes.

Starting from the Gaussian free theory, the natural functional measure is:
$$
[d\phi]\,=\,\prod_j d\left(\f{\phi^j}{\sqrt{d_j}}\right)=\prod_j d(\sqrt{d_j}\,M_j).
$$
Defining the renormalized matrices $\tl{M}_j=\sqrt{d_j}\,M_j$ allows to reabsorb the $d_j$ factor
into the potential:
\be
Z\,=\,\prod_j \int[d\tl{M}_j]\, e^{-\tr V_j[\tl{M}_j]},
\ee
where the potentials $V_j$ now depend on the representation $j$ (thus on the matrix size) but have
a trivial quadratic term:
\be
V_j[\tl{M}]\,=\,
\f12 \tl{M}^2+\sum_{n\ge 3} \f{\alpha_n}{n!(\sqrt{d_j})^{n-2}}\tl{M}^n.
\ee
The $d_j$ factors only disappear completely when the potential is purely quadratic, i.e when we
consider only the mass term in the group field theory and no interaction term. Otherwise $d_j$
factors come into the matrix couplings. The orthogonal polynomials involved in the exact
computation of the matrix partition function are different for different matrix sizes and the $h_n$
factors now also depend on the representation $j$. To solve exactly the group field theory, we need
to compute this whole family of matrix models with rescaled potentials. A last remark is that as
$j$ grows to infinity, the interaction couplings are sent to 0 and the potential $V_j$ becomes
almost purely quadratic.

The natural issue to investigate is the existence of critical couplings. This question has been
well-studied and solved as far as matrix models are concerned. For instance, in the simplest case
of a quartic potential:
$$
V[M]=\f12M^2+\f{\alpha}{d_j}M^4,
$$
we know that the critical coupling in the large matrix size limit $d_j\arr\infty$ is
$\alpha_c=-\f1{48}$. It would be interesting to check whether the tower of matrix models affects
this result or if only the behavior for large matrix size matters. If large sizes dominate the
partition function, then the results on the $1/N$ expansion of matrix models should be enough to
understand the behavior of the 2d GFT. Otherwise, we would need to take into account the
contribution of small matrix sizes.

\medskip

We now turn to the generic class of 2d group field theory with non-trivial kinetic term and
invariant under the action of the deformed Poincar\'e group:
$$
S[\phi]\,=\,\f12\int dg\,\phi(g)\kk(g)\phi(g^{-1}) + \sum_{n\ge
3}\f{\alpha_n}{n!}\int[dg]^n\,\delta(g_1..g_n)\phi(g_1)..\phi(g_n).
$$
We can again write such a field theory in term of matrices by decomposing the field in $\SU(2)$
representations. However, as soon as $\kk(g)$ is non-constant on the group, the quadratic term
couples matrices of different sizes and the partition function can not be formulated to a tower of
uncoupled matrix models \cite{phase}. We focus on the special case $\kk(g)=\chi_{1/2}(g)/ 2+
c\,=\cos\theta+c$ where $c$ is an arbitrary (real) number and restrict the interactions to the
single $n=4$ term to keep notations simple. The action can then be written as:
\be
S\,=\,\f14\sum_{j,k}\phi^j_{ab}\phi^k_{\tilde{b}\tilde{a}}
\left(\begin{array}{cc|c}
j &\f12 & k \\
a &m & \tilde{a}
\end{array}\right)
\left(\begin{array}{c|cc}
k& j &\f12 \\
\tilde{b}& b &m
\end{array}\right)
{}+\f{c}2\sum_j\f{1}{d_j}\tr(\phi^j)^2 +\f{\alpha}{4!}\sum_j\f{1}{d_j^3}\tr(\phi^j)^4,
\ee
where the new kinetic term is evaluated using the Clebsh-Gordan coefficients. The important
property of such actions is that the coupling between matrix sizes induced by the Clebsh-Gordan
coefficients breaks the invariance under unitary matrices. Therefore, we can not act independently
by unitaries on each matrix $\phi^j$ and we do not gauge-fix by diagonalizing the matrices.
Nevertheless, we still have a theory invariant under the deformed Poincar\'e group, but we can not
deal with each matrix size separately like the case studied above.

Starting from the standard approach to matrix model, the simplest way to compute correlations would be to consider the mixing term $\sum_{j,k}\phi^j\phi^k$ as an interaction term and expand it:
\be
Z= \int [d\phi]\, e^{iS[\phi]}\,=\,
\sum_K \f1{K!}\int [d\phi]\,
\left(\f14\sum_{j,k}\phi^j_{ab}\phi^k_{\tilde{b}\tilde{a}}\right)^K
e^{\f{c}2\sum_j\f{1}{d_j}\tr(\phi^j)^2 +\f{\alpha}{4!}\sum_j\f{1}{d_j^3}\tr(\phi^j)^4}.
\ee
Since we keep in the action only the term that do not couple the different matrix sizes, we could them directly use the already-known calculations. Then one would check whether the sum over $K$ is convergent or not. For $K=0$, we have the standard uncoupled matrix model partition function $Z_{uncoupled}$. The $K=1$ term vanishes by parity. The next term $K=2$ involves the product of correlations $\la(\phi^j)^2\ra_{uncoupled}\,\la(\phi^k)^2\ra_{uncoupled}$ times a product of four Clebsh-Gordan coefficients. This actually leads to a $\{6j\}$-symbol. This is hardly surprising, since the 2d GFT is understood to be a sector of the 3d GFT whose Feynmann diagrams are given by the Ponzano-Regge spinfoam amplitudes i.e some products of $\{6j\}$-symbols. It would be interesting to see if we can truly compute all the term for arbitrary $K$.

It seems this would lead to non-trivial results from the point of view of the non-commutative field theory. Indeed, we could compute the propagator $\la \phi(g)\phi(h)\ra$ as a sum of $\la \phi^j \phi^k\ra$ correlations. It would be very interesting to see if the sum over $K$ can be given a non-perturbative meaning.

We also propose a different approach: integrating the matrix modes one per one starting from the $j=0$
mode as in a renormalisation group calculation. Starting with $j=0$, the $\phi^0$ mode is only
coupled to the $\phi^{1/2}$ mode due to our special choice of $\kk(g)$. Focusing on the terms
involving $\phi^0$, the partition function reads:
\be
Z\,=\,\int \prod_j[d\phi^j]\,
e^{-\f{c}{2}(\phi^0)^2-\f{\alpha}{3!}(\phi^0)^3-\f14\phi^0\tr\phi^{1/2} -\f{c}4\tr(\phi^{1/2})^2
-\dots}
\ee
Keeping in mind that $\phi^0$ is a single real number, we can integrate over it when we set the
interaction coupling to $\alpha=0$. This gives:
\be
Z\,=\,\int \prod_{j>0}[d\phi^j]\, \sqrt{\f{2\pi}{c}}\,e^{\f{1}{32c}\left(\tr\phi^{1/2}\right)^2
-\f{c}4\tr(\phi^{1/2})^2 -\dots}
\ee
The $j=1/2$ mode will then only couple to $\phi^1$ and we could move step by step to higher $j$'s
to compute the partition function only dealing with Gaussian integrals.  In order to take the
interaction into account, we would expand the partition as usual in powers of the coupling
$\alpha$.

It is likely that techniques developed for matrix model could help to evaluate this partition function and see if there exists a critical regime. Moreover, computing this multi-matrix model would most likely also help understanding the physical properties of the associated non-commutative quantum field theory with the deformed Poincar\'e invariance.

\section*{Conclusion}

Group field theory turned out to be a very useful tool to formalize spinfoam models. They are shown
to be closely related to matrix models. And they appeared to be also related to non-commutative
field theories through the recently developed group Fourier transform. I would like to insist on
two points.

First, group field theories for spinfoam models are non-commutative field theories. This statement
can be used both ways. We can use our knowledge of spinfoam models and the relation between the
group field theories and topological field theories to study some specific examples of
non-commutative QFTs. But we should also be aware that the issues encountered when studying
non-commutative field theories will occur in the study of group field theory at some stage, among
which defining the right propagator, infrared-ultraviolet mixing in the renormalisation process and
ambiguities in the statistics and path integral measure.

Second, the framework presented in this paper allows to make a direct link between matrix models
and non-commutative geometry. This context opens the door to a constructive exchange of tools. On
the one hand, matrix models are integrable systems and we could use these methods to probe the
structure of group field theories and solve some models of non-commutative field theories. On the
other hand, we showed that the non-commutative fields theories have a non-trivial Poincar\'e
invariance and that they lead to new matrix models with non-trivial couplings between matrices of
different sizes: it would be interesting to see whether these matrix models are also integrable or
not and if this deformed Poincar\'e symmetry is relevant to the structure of matrix models.

%
%



\begin{thebibliography}{99}





\bibitem{2d}
E.R. Livine, A. Perez and C. Rovelli,
{\it 2d manifold-independent spinfoam theory},
Class. Quant. Grav. [arXiv:gr-qc/0102051]; \\
A. Mikovic,
{\it Quantum Field Theory of Spin Networks},
Class.Quant.Grav. 18 (2001) 2827-2850 [arXiv:gr-qc/0102110]

\bibitem{boulatov}
D. Boulatov,
{\it A Model of Three-Dimensional Lattice Gravity},
Mod.Phys.Lett. A7 (1992) 1629-1646 [arXiv:hep-th/9202074]

\bibitem{gft}
M.P. Reisenberger, C. Rovelli,
{\it Spacetime as a Feynman diagram: the connection formulation},
Class.Quant.Grav. 18 (2001) 121-140 [arXiv:gr-qc/0002095]

\bibitem{laurent}
L. Freidel,
{\it Group Field Theory: An overview},
Int.J.Theor.Phys. 44 (2005) 1769-1783 [arXiv:hep-th/0505016]

\bibitem{phase}
W.~J.~Fairbairn and E.~R.~Livine,
  {\it 3d spinfoam quantum gravity: Matter as a phase of the group field theory,}
  Class.\ Quant.\ Grav.\  {\bf 24}, 5277 (2007)
  [arXiv:gr-qc/0702125].

\bibitem{PR3}
L. Freidel and E.R. Livine,
{\it Ponzano-Regge model revisited III: Feynman diagrams and Effective field theory},
Class.Quant.Grav. 23 (2006) 2021-2062 [arXiv:hep-th/0502106].


\bibitem{effqg}
L. Freidel and E.R. Livine,
{\it 3d Quantum Gravity and Effective Non-Commutative Quantum Field Theory},
Phys.Rev.Lett. 96 (2006) 221301 [arXiv:hep-th/0512113].

\bibitem{4dgftdsr}
F. Girelli, E.R. Livine and D. Oriti,
{\it Doubly Special Relativity from 4d Spinfoam models}, in preparation; \\
E.R. Livine, {\it Non-commutative field theories from 3d and 4d spin foam models}, Talk at the
``Noncommutative Deformations of Special Relativity" ICMS workshop (Edinburgh, July 2008)

\bibitem{catherine}
C. Meusburger,
{\it Quantum double and $\kappa$-Poincaré symmetries in (2+1)-gravity and Chern-Simons theory},
arXiv:0809.0052; \\
C. Meusburger and B. Schroers,
{\it Generalised Chern-Simons actions for 3d gravity and kappa-Poincare symmetry},
Nucl.Phys.B806 (2009) 462-488 [arXiv:0805.3318]


\bibitem{2dencore}
D. Oriti, C. Rovelli and S. Speziale,
{\it Spinfoam 2d quantum gravity and discrete bundles},
Class.Quant.Grav. 22 (2005) 85-108 [arXiv:gr-qc/0406063]

\bibitem{majid}
L. Freidel and S. Majid,
{\it Noncommutative Harmonic Analysis, Sampling Theory and the Duflo Map in 2+1 Quantum Gravity},
arXiv:hep-th/0601004

\bibitem{karim}
E. Joung, J. Mourad and K. Noui,
{\it Three Dimensional Quantum Geometry and Deformed Poincare Symmetry},
arXiv:0806.4121

\bibitem{PR1}
L. Freidel and D. Louapre,
{\it Ponzano-Regge model revisited I: Gauge fixing, observables and interacting spinning particles},
Class.Quant.Grav. 21 (2004) 5685-5726 [arXiv:hep-th/0401076]

\bibitem{jimmy}
E.R. Livine and J.P. Ryan,
{\it A Note on B-observables in Ponzano-Regge 3d Quantum Gravity},
arXiv:0808.0025

\bibitem{bh}
E.R. Livine and D.R. Terno,
{\it  Quantum Black Holes: Entropy and Entanglement on the Horizon},
Nucl.Phys. B741 (2006) 131-161 [arXiv:gr-qc/0508085]

\bibitem{4ddsr}
L. Freidel, J. Kowalski-Glikman, S. Nowak,
{\it From noncommutative kappa-Minkowski to Minkowski space-time},
arXiv:hep-th/0612170

\bibitem{withflo}
F. Girelli and E.R. Livine,
{\it Group Field Theory: Classical Solutions, Poincar\'e Invariance and Permutations},
in preparation

\bibitem{borelsum}
L. Freidel and D. Louapre,
{\it  Non-perturbative summation over 3D discrete topologies},
Phys.Rev. D68 (2003) 104004  [arXiv:hep-th/0211026]

\bibitem{madore}
J. Madore,
{\it The fuzzy sphere},
Class. Quant. Grav. 9 (1992) 69-87

\bibitem{fuzzy}
L. Freidel and K. Krasnov,
{\it The Fuzzy Sphere $\star$-product and Spin Networks},
J.Math.Phys. 43 (2002) 1737-1754 [arXiv:hep-th/0103070]

\bibitem{WZW}
A.Yu. Alekseev, A. Recknagel and V. Schomerus,
{\it  Non-commutative World-volume Geometries: Branes on SU(2) and Fuzzy Spheres},
JHEP 9909 (1999) 023 [arXiv:hep-th/9908040]

\bibitem{coherent}
E.R. Livine and D. Oriti,
{\it Coherent States for 3d Deformed Special Relativity: semi-classical points in a quantum flat
spacetime}, JHEP 0511 (2005) 050 [arXiv:hep-th/0509192]; \\
E.R. Livine and S. Speziale,
{\it A new spinfoam vertex for quantum gravity},
Phys. Rev. D 76, 084028 (2007) [arXiv:0705.0674]

\bibitem{matrix}
M. Mari\~{n}o,
{\it Nonperturbative effects and nonperturbative definitions in matrix models and topological
strings}, arXiv:0805.3033; \\
A. Morozov,
{\it Matrix Models as Integrable Systems},
arXiv:hep-th/9502091

\end{thebibliography}
\end{document}